\documentclass[aps, prd, twocolumn, lengthcheck, superscriptaddress, 
nofootinbib]{revtex4-1}
\usepackage[utf8]{inputenc}
\usepackage[sort&compress]{natbib}
\usepackage{ulem}
\usepackage{bm}
\usepackage{times}
\usepackage{amssymb,amsbsy,amsmath,amsfonts}
\usepackage{graphicx}
\usepackage{float}
\usepackage{color}
\usepackage{morefloats}
\usepackage{rotating}
\usepackage{srcltx}
\usepackage{overpic}
\usepackage{slashed}
\usepackage{subfigure}
\usepackage{multirow}
\usepackage{verbatim}
\usepackage{hyperref}
\usepackage{tabularx}

\begin{document}
\title{Bridging doubly heavy tetraquark mass spectrum with heavy baryons utilizing heavy antiquark-diquark symmetry}

\author{Liu-Yu Zhang}
\affiliation{School of Science, Shenzhen Campus of Sun Yat-sen University, Shenzhen 518107, China}

\author{Tian-Wei Wu}
\email{wutw6@mail.sysu.edu.cn}
\affiliation{School of Science, Shenzhen Campus of Sun Yat-sen University, Shenzhen 518107, China}

\author{Yong-Liang Ma}
\email{ylma@nju.edu.cn}
\affiliation{School of Frontier Sciences, Nanjing University, Suzhou, 215163, China}

\begin{abstract}
Motivated by the observation of the doubly charmed tetraquark $T_{cc}(3875)^+$, we present a systematic study of doubly heavy tetraquarks ($T_{QQ'\bar{q}\bar{q}'}$) using heavy antiquark-diquark symmetry (HADS) within a constituent quark model. 
By calibrating model parameters to known hadron spectra and incorporating the effective mass formula, we predict the masses for 38 ground-state tetraquarks with $cc$, $bb$, and $bc$ heavy quark pairs, including the non-strange, single-strange, and double-strange configurations with quantum numbers $J^P = 0^+, 1^+$ and $2^+$. 
Notably, we identify several stable states below the relevant meson-meson thresholds, particularly in the $bb\bar{q}\bar{q}'$ sector. The explicit connection between doubly heavy tetraquark and heavy baryon spectra through HADS reduces model dependence and reveals fundamental systematics in the heavy-quark hadron landscape.

\end{abstract}
\maketitle

\section{Introduction}

The discovery of exotic hadrons beyond the conventional quark-antiquark (meson) and three-quark (baryon) paradigms stands as a pivotal achievement of modern particle physics. Since 2003, many exotic hadronic states which cannot be well understood by the conventional constituent quark model have been observed. Typically, in the heavy flavor sector, the observations of the so-called $X, Y, Z$ states, the tetraquarks, the pentaquarks and so on, have extended the hadron spectra and deepen our understanding on the strong interactions between quarks in the nonperturbative region (see, e.g., Refs.~\cite{Hosaka:2016pey,Ali:2017jda,Guo:2017jvc,Esposito:2016noz,Lebed:2016hpi,Richard:2016eis,Chen:2016qju,Liu:2019zoy,Brambilla:2019esw,Liu:2024uxn,Wang:2025sic} for reviews). 

In 2017, the doubly heavy baryon $\Xi_{cc}^{++}$ was discovered in the $\Lambda_c^+K^-\pi^+\pi^+$ mass spectrum by the LHCb collaboration with a mass $3621.4$ MeV~\cite{LHCb:2017iph}, which is the first observed doubly heavy hadron ever.
In 2021, the observation of the doubly charmed tetraquark candidate $T_{cc}^+(3875)$ by the LHCb collaboration marks a highlight moment~\cite{LHCb:2021vvq,LHCb:2021auc}. Its striking proximity to the $D^{*+}D^0$ mass threshold (within $\sim$360 keV) and narrow width ignited intense theoretical debate concerning its fundamental nature. Two primary interpretations emerged: a loosely bound hadronic molecule held together by residual strong forces between $D^{*}$ and $D$ mesons~\cite{Du:2021zzh,Padmanath:2022cvl,Feijoo:2021ppq,Meng:2021jnw,Deng:2021gnb}, and a compact tetraquark state with internal structure dominated by strong, short-distance correlations within the $c\bar{c}u\bar{d}$ valence quarks~\cite{Karliner:2017qjm,Agaev:2021vur,Kim:2022mpa,Wu:2022gie}. 
All in all, studies so far have not determined the internal structure and nature of $T_{cc}(3875)^+$, the doubly heavy tetraquarks needs more studies to reveal its nature. 
While both pictures can qualitatively accommodate the mass and quantum numbers of the $T_{cc}^+(3875)$, they predict significantly different properties for its partners and the broader spectrum of doubly heavy tetraquarks (DHTs). Unambiguously resolving this structural dichotomy remains a central challenge and arouse the interests of studies on the doubly heavy hadrons both theoretically and experimentally~\cite{Karliner:2017qjm,Eichten:2017ffp,Wang:2017dtg,Wang:2017uld,Cheng:2020wxa,Chen:2021vhg,Ren:2021dsi,Deng:2021gnb,Xin:2021wcr,Albaladejo:2021vln,Ling:2021bir,Meng:2021jnw,Fleming:2021wmk,Feijoo:2021ppq,Guo:2021yws,Du:2021zzh,Huang:2021urd,Wu:2021kbu,Luo:2021ggs,Abreu:2022sra,Yan:2021wdl,Padmanath:2022cvl,Agaev:2021vur,Kim:2022mpa,Wu:2022gie,Ebert:2007rn,Luo:2017eub,Lu:2020rog,Zhang:2021yul,Weng:2021hje,Mutuk:2023oyz,Ma:2023int,Liu:2025jyn}.

Current theoretical efforts to map the DHTs landscape employ diverse methodologies. In Ref.~\cite{Karliner:2017qjm}, the authors have successfully predicted the existence of the doubly heavy tetraquarks before the discovery of $T_{cc}(3875)^+$ employing the constituent quark model with a good light quark symmetry. 
The masses of tetraquarks  were calculated in the framework of the diquark-antidiquark picture in the relativistic quark model~\cite{Ebert:2007rn}, in a relativized quark model within the variational method~\cite{Lu:2020rog}, and an extended chromomagnetic model~\cite{Weng:2021hje}. 
Ref.~\cite{Luo:2017eub,Guo:2021yws} systematically studied the mass splittings of the DHT states with the color-magnetic interaction by considering color mixing effects and estimate roughly their masses.  
 The masses and magnetic moments of tetraquarks  were computed in a unified framework of the MIT bag model~\cite{Zhang:2021yul} and diffusion Monte Carlo method (DMC)~\cite{Mutuk:2023oyz}. 
 In Ref.~\cite{Ma:2023int}, the authors used DMC to calculate the DHT system in two kinds of constituent quark models, the pure constituent quark model AL1/AP1 and the chiral constituent quark model.

Phenomenological Quark Models offer intuitive structural insights, yet conventional treatments often yield masses systematically higher than experimental candidates, suggesting the necessity to incorporate crucial dynamical correlations.
A powerful organizing principle emerges from heavy-quark symmetry. In the heavy quark mass limit ($m_Q \to \infty$, $Q=c,b$), the spin of the heavy quark decouples and heavy diquark shares the same color structure as  heavy antiquark, giving rise to heavy antiquark-diquark symmetry~\cite{Savage:1990di,Georgi:1990ak,Carone:1990pv} (HADS, also known as super flavor symmetry~\cite{Ma:2015lba,Ma:2015cfa,Ma:2017nik}). HADS establishes an approximate symmetry between the color-antitriplet heavy diquark ($QQ$) in a doubly heavy baryon ($QQ'q$, $q=u,d,s$) and the color-triplet heavy antiquark ($\bar{Q}$) in a heavy-light meson ($\bar{Q}q$). Consequently, the spectrum of doubly heavy baryons ($\Xi_{cc}$, $\Xi_{bb}$, $\Omega_{cc}$, etc.) and heavy-light mesons ($D^{(*)}$, $B^{(*)}$, $D_s^{(*)}$, etc.) exhibit remarkable parallels dictated by light degrees of freedom~\cite{Hu:2005gf,Brambilla:2005yk,Ma:2017nik}. Crucially, this symmetry extends to compact DHTs ($QQ'\bar{q}\bar{q}'$), where the $QQ'$ diquark core plays a role analogous to the single heavy antiquark $\bar{Q}$, while the light antidiquark ($\bar{q}\bar{q}'$) remains unchanged. The $QQ'\bar{q}\bar{q}'$ tetraquark spectrum should therefore reflect symmetries connecting the $\bar{Q}\bar{q}\bar{q}'$ baryon spectrum.

In this work, we leverage the profound implications of HADS within a robust constituent quark model (CQM) framework to systematically investigate the mass spectrum of DHT states. Our approach transcends the limitations of treating tetraquarks in isolation by explicitly incorporating the symmetry connecting compact tetraquarks to the well-established spectra of heavy baryons and  mesons. This imposes stringent constraints on model dynamics and predicted masses. We employ a sophisticated CQM incorporating effective mass formula calibrated on established hadron spectra~\cite{Karliner:2008sv,Karliner:2017qhf,Karliner:2014gca,Wu:2022gie}. Within this framework, we construct compact tetraquark configurations where the $QQ'$ pair forms a colored antitriplet diquark, and the light $\bar{q}\bar{q}'$ pair forms a color triplet, combining to color-singlet states. 
Our study delivers comprehensive predictions for ground states of $cc\bar{q}\bar{q}'$, $bb\bar{q}\bar{q}'$, and $bc\bar{q}\bar{q}'$ systems with various light quark flavors and quantum numbers. We critically compare predictions for $T_{cc}$ and its partners with experimental data and other theoretical approaches. Key spectral features—state ordering, hyperfine splittings, and proximity to meson-meson thresholds—serve as discriminants between compact tetraquark and molecular interpretations. Furthermore, we predict masses and quantum numbers for unobserved states, providing essential guidance for experimental searches at LHCb, Belle II, and PANDA.

This paper is structured as follows: Sect.~\ref{Frame} introduces the doubly heavy tetraquark configurations within HADS, the constituent quark model and the mass formulas. Sect.~\ref{Results} presents the predicted mass spectra with the corresponding meson-meson thresholds and a comparison with other studies. A summary is given in the last section.

\section{Theoretical framework}
\label{Frame}

In this section, we will first introduce the HADS that correlates the DHTs with heavy baryons. Subsequently, we will elaborate on how to utilize HADS to construct the configurations of DHT states that are correlated with heavy baryons. Finally, we will explain how to calculate the mass spectra of these doubly heavy tetraquarks using the mass formulas derived from the constituent quark model.

HADS is a significant concept in the study of hadronic physics~\cite{Savage:1990di,Georgi:1990ak,Carone:1990pv,Wu:2020rdg}. In the limit of heavy quark masses approaching infinity in Quantum Chromodynamics (QCD), this symmetry emerges, connecting the properties of hadrons with two heavy quarks $QQ'$ to those with one heavy antiquark $\bar{Q}$. Physically, HADS arises when a diquark pair forms a tightly bound, nearly point-like object. For instance, the attraction between two heavy quarks in a diquark comes from a color Coulombic interaction in the color $3_c$ channel. When the quark masses are large enough, the heavy quarks move slowly, acting like non-relativistic particle sources in a Coulombic potential. As the size of a Coulombic bound state is inversely proportional to its mass (for a fixed coupling), in the large mass limit, the diquark becomes a heavy, small object with color $\bar{3}_c$, effectively acting as a static point-like 
$\bar{3}_c$ color source, similar to a heavy antiquark.

This symmetry has important implications for understanding the spectra and properties of hadrons. It relates the quark structure of doubly heavy baryons to that of heavy antimesons. For example, the masses of $\Xi_{cc}^{(*)}$ baryons and $D^{(*)}$ mesons serves the relation $m_{\Xi^*_{cc}}-m_{\Xi_{cc}}=\frac{3}{4}(m_{\bar{D}^*}-m_{\bar{D}})$ has been numerically confirmed with the masses observed by experiments ($m_{\Xi_{cc}}$, $m_{\bar{D}^*}$ and $m_{\bar{D}}$) and lattice QCD simulations ($m_{\Xi_{cc}^*}$)~\cite{Padmanath:2015jea,Chen:2017kxr,Alexandrou:2017xwd,Mathur:2018rwu}.
Along this line, the masses and configurations of DHTs can also be related to those of heavy baryons by leveraging HADS, which relates DHTs to heavy baryons by treating the double-heavy diquark ($QQ'$) as a heavy antiquark $\bar{X}$.  Combined with the constituent quark model—well-established for describing meson and baryon spectra—this symmetry allows us to predict the mass spectrum of DHTs systematically. We extend previous studies to include not only doubly charmed ($T_{cc}$) and doubly bottomed ($T_{bb}$) states but also bottom-charmed ($T_{bc}$) configurations, covering non-strange, single-strange, and double-strange light quark combinations. The predicted states span spin-parity quantum numbers 
$0^+$, $1^+$ and $2^+$, providing a complete landscape of ground states of DHTs.

\subsection{Configurations of DHTs}
The configurations of DHTs are labeled with $[[\bar{q}\bar{q}']^{c}_{s}[QQ']^{C}_{S}]^I_J$. Where $C$ and $S$ is the color and spin of the heavy diquark $QQ'$, respectively. $c$ and $s$ is the color and spin of the light diquark $\bar{q}\bar{q}'$, respectively.  $I$ is the isospin of the tetraquark and $J$ is the total angular momentum of the tetraquark. DHTs with different number of strange quarks have  different configurations. 
Since we are establishing a connection between DHTs and heavy baryons, we assign $c$
 and $C$ to be $3$ and $\bar{3}$, respectively. Color sextet combinations may also exist in the color structure of 
DHT states. However, such configurations have not been experimentally observed to date. In diquarks, the Casimir coefficient for the $\mathbf{3}_c \otimes \mathbf{3}_c = \mathbf{6}_c$ representation is $1/3$, indicating a repulsive interaction between the two quarks in this color configuration, while the coefficient for $\mathbf{3}_c \otimes \mathbf{3}_c = \bar{\mathbf{3}}_c$ is $-2/3$, suggesting a strong attractive interaction. This implies that $\mathbf{3}_c \otimes \mathbf{3}_c = \bar{\mathbf{3}}_c$ is likely the dominant form of diquark coupling. Studies in Ref.~\cite{Lu:2020rog} show that the weight of the latter configuration in DHT states exceeds 90\%. Based on these theoretical and experimental reasons, in this work, we consider only the $\mathbf{3}_c$ and $\bar{\mathbf{3}}_c$ color configurations for diquarks. For $s$ and $S$, their values can be either 0 or 1, depending on the flavor combination of the tetraquark. The configurations are listed below with three combinations on order of numbers of strange quarks, unflavored light quarks are denoted as $n$ with $n=u,d$.

1. DHTs with non-strange light quarks

(a) Configuration $[[\bar{n}\bar{n}]^{3}_{s}[QQ]^{\bar{3}}_{1}]^I_J$:
\begin{equation*}
    \begin{split}
&[[\bar{n}\bar{n}]^3_0[QQ]^{\bar{3}}_1]^0_1\\
&[[\bar{n}\bar{n}]^3_1[QQ]^{\bar{3}}_1]^1_0\\
&[[\bar{n}\bar{n}]^3_1[QQ]^{\bar{3}}_1]^1_1\\
&[[\bar{n}\bar{n}]^3_1[QQ]^{\bar{3}}_1]^1_2\\
\end{split}
\end{equation*}

(b) Configuration $[[\bar{n}\bar{n}]^{3}_{s}[cb]^{\bar{3}}_{S}]^I_J$:
\begin{equation*}
    \begin{split}
&[[\bar{n}\bar{n}]^3_0[bc]^{\bar{3}}_0]^0_0\\
&[[\bar{n}\bar{n}]^3_0[bc]^{\bar{3}}_1]^0_1\\
&[[\bar{n}\bar{n}]^3_1[bc]^{\bar{3}}_0]^1_1\\
&[[\bar{n}\bar{n}]^3_1[bc]^{\bar{3}}_1]^1_0\\
&[[\bar{n}\bar{n}]^3_1[bc]^{\bar{3}}_1]^1_1\\
&[[\bar{n}\bar{n}]^3_1[bc]^{\bar{3}}_1]^1_2\\
    \end{split}
\end{equation*}

2. DHTs with single-strange light quarks

(a) Configuration $[[\bar{n}\bar{s}]^{3}_{s}[QQ]^{\bar{3}}_{1}]^{\frac{1}{2}}_J$:
\begin{equation*}
    \begin{split}
&[[\bar{n}\bar{s}]^3_0[QQ]^{\bar{3}}_1]^{\frac{1}{2}}_1\\
&[[\bar{n}\bar{s}]^3_1[QQ]^{\bar{3}}_1]^{\frac{1}{2}}_0\\
&[[\bar{n}\bar{s}]^3_1[QQ]^{\bar{3}}_1]^{\frac{1}{2}}_1\\
&[[\bar{n}\bar{s}]^3_1[QQ]^{\bar{3}}_1]^{\frac{1}{2}}_2\\
    \end{split}
\end{equation*}

(b) Configuration $[[\bar{n}\bar{s}]^{3}_{s}[bc]^{\bar{3}}_{S}]^{\frac{1}{2}}_J$:
\begin{equation*}
    \begin{split}
&[[\bar{n}\bar{s}]^3_0[bc]^{\bar{3}}_0]^{\frac{1}{2}}_0\\
&[[\bar{n}\bar{s}]^3_0[bc]^{\bar{3}}_1]^{\frac{1}{2}}_1\\
&[[\bar{n}\bar{s}]^3_1[bc]^{\bar{3}}_0]^{\frac{1}{2}}_1\\
&[[\bar{n}\bar{s}]^3_1[bc]^{\bar{3}}_1]^{\frac{1}{2}}_0\\
&[[\bar{n}\bar{s}]^3_1[bc]^{\bar{3}}_1]^{\frac{1}{2}}_1\\
&[[\bar{n}\bar{s}]^3_1[bc]^{\bar{3}}_1]^{\frac{1}{2}}_2\\
     \end{split}
\end{equation*}

3. DHTs with double-strange quarks:

(a) Configuration $[[\bar{s}\bar{s}]^{3}_{1}[QQ]^{\bar{3}}_{1}]_J$:
\begin{equation*}
    \begin{split}
&[[\bar{s}\bar{s}]^3_1[QQ]^{\bar{3}}_1]_0\\
&[[\bar{s}\bar{s}]^3_1[QQ]^{\bar{3}}_1]_1\\
&[[\bar{s}\bar{s}]^3_1[QQ]^{\bar{3}}_1]_2\\
    \end{split}
\end{equation*}

(b) Configuration $[[\bar{s}\bar{s}]^{3}_{1}[bc]^{\bar{3}}_{S}]_J$:
\begin{equation*}
    \begin{split}
&[[\bar{s}\bar{s}]^3_1[bc]^{\bar{3}}_0]_1\\
&[[\bar{s}\bar{s}]^3_1[bc]^{\bar{3}}_1]_0\\
&[[\bar{s}\bar{s}]^3_1[bc]^{\bar{3}}_1]_1\\
&[[\bar{s}\bar{s}]^3_1[bc]^{\bar{3}}_1]_2\\
    \end{split}
\end{equation*}

\begin{table}[htbp]
	\caption{Masses, bindings, and spin hyperfine splittings of baryons in CQM (in unit of MeV).}
    \label{Paras: Baryon}
	\centering
		\begin{tabular}{ccccccc}
         \hline\hline
Parameters&$m_n$&$m_s$&$m_c$&$m_b$&$B_{cs}$&$B_{bs}$\\
Values&364.3 &536.2 &1715.9 &5047.3 &53.4 &62.6 \\
\hline
$B_{cc}$&$B_{cb}$&$B_{bb}$ &$\alpha/{m_n^2}$
&$\alpha_{cc}/m_c^2$&$\alpha_{cb}/m_cm_b$&$\alpha_{bb}/m_b^2$\\
217.7 & 256.2  & 422.0 &$-76.8$&$-21.2$ &$-12.7$ &$-11.6$ \\
			\hline\hline
			\end{tabular}
	\end{table}

\begin{table}[htbp]
	\caption{Masses, bindings, and hyperfine couplings of quarks in doubly heavy tetraquark ground states (in unit of MeV).}
    \label{Paras:DHT}
	\centering
		\begin{tabular}{ccccc}
         \hline\hline
Parameters&$m_{[cc]^{\bar{3}}_1}$&$m_{[bb]^{\bar{3}}_1}$&$m_{[cb]^{\bar{3}}_0}$&$m_{[cb]^{\bar{3}}_1}$\\
Values&3300.8 &9821.0 &6567.0 &6600.9   \\
            \hline
Parameters&$B_{[cc]s}$&$B_{[bb]s}$&$B_{[cb]s}$& $\alpha/m_nm_{[cc]_{1}^{\bar{3}}}$\\
Values&106.8 &125.2 &116.0& $-8.5$\\
			\hline\hline
			\end{tabular}
	\end{table}

\subsection{Mass formula of DHT in CQM}

 In this subsection, we use a well-established mass formula in CQM which have been proved successfully interpret the spectra of mesons and baryons~\cite{Karliner:2008sv,Karliner:2014gca,Wu:2022gie}, to calculate the masses of DHTs with HADS.
 In Ref.~\cite{Karliner:2017qjm},  this model is used to study the doubly heavy tetraquarks with a good light quark symmetry and successfully predict the mass of $T_{cc}$(3875).
 In Ref.~\cite{Wu:2022gie}, the ground states of heavy baryon spectrum is well described with the mass formula
\begin{equation} 
M_{B}=\sum_{i}m_i + \sum_{i<j}(F_i\cdot F_j)[B_{ij}+(\sigma_i\cdot\sigma_j)\alpha_{ij}/m_im_j]
\end{equation}
where $m_i, B_{ij}$ and $\alpha_{ij}$ are the mass of the $i$th constituent quark, the binding and hyperfine spin coupling constant between the $i$th and $j$th quarks, respectively. $F_i, \sigma_i$ are the color and spin operators with
$i, j = 1, 2, 3$, respectively. This simple formula yields a standard mass variance $\chi_{M}=7.6$ MeV. The method is successfully extended to doubly charmed and doubly bottomed tetraquarks with non-strange light quarks~\cite{Wu:2022gie}. Here in this work, besides what is studied before, we extend the study to bottom-charmed sector and cases that with non-, single-, and double-strange quark(s). Thus, the complete mass spectrum of the whole ground states of DHTs correlated to the heavy baryons are displayed.

The mass formula in CQM can be extended to DHT with HADS, reads
\begin{equation}
\label{MDHT}
    M_{DHT}=m_{[QQ']^{C}_Sqq'} + \sum_{i<j}(F_i\cdot F_j)[B_{ij}+(\sigma_i\cdot\sigma_j)\alpha_{ij}/m_im_j],
\end{equation}
where
$m_{[QQ']^{C}_{S}qq'}=m_{[QQ']^{\bar{3}}_{S}}+m_q+m_{q'}$ is the constituent mass of the quark components with $m_{[QQ']^{\bar{3}}_{S}}$ and $m_{q^{(\prime)}}$ being, respectively, the effective masses of heavy diquark and light quark. The coefficients of color structure operators are determined by the color representations of the quark components and the group theory of 
$SU(3)_c$, where is $-2/3$ for $\bar{3}_c$ representation between two quarks in baryons.  
$B_{ij}$ and $\alpha_{ij}$ are the bindings and spin hyperfine coupling constants between the light quarks and those between the diquark $[QQ']^C_S$ and the light quark. 
The effective mass of the heavy diquark $m_{[QQ']^{C}_S}$ can also be determined by the mass formula in CQM
\begin{equation}
m_{[QQ']^{C}_S}=m_{Q}+m_{Q'}+F_Q\cdot F_{Q'}[B_{QQ'}+\frac{\alpha _{QQ'}\sigma_Q\cdot\sigma_{Q'}}{m_{Q}m_{Q'}}],
\end{equation}
where $m_{Q^{(\prime)}}$ is the heavy quark mass.
The quark masses, bindings and spin hyperfine coupling constants are needed to calculate the mass of the heavy diquark $[QQ']^{C}_S$.

All the required parameters are listed in Table~\ref{Paras: Baryon}, of which the values are determined by fitting the baryon spectra and corresponding hidden-flavor and bottom-charmed heavy mesons~\cite{Wu:2022gie,Karliner:2014gca}. It should be noted that the masses of constituent quarks in baryons are diffrent with those in mesons. In this work, since quarks are in the form of diquarks, the quark masses we adopt for calculating the mass spectra are all based on the quark masses within baryons.  Only when determining the hyperfine splitting constants $\alpha_{QQ'}$ and binding energies $B_{QQ'}$ between heavy quarks, given that the corresponding heavy baryons have not yet been experimentally observed, we resort to meson masses for such determinations. Take $bb$ as example, since $m_{\Upsilon _b}=2m_b^m+(-4/3)[B^m_{bb}+\alpha _{bb}/(m_b^m)^2]$ and $m_{\eta _{b}}=2m_b^m+(-4/3)[B^m_{bb}-3\alpha _{bb}/(m_b^m)^2]$, we have $\alpha_{bb}/(m_b^b)^2\approx\alpha_{bb}/(m_b^m)^2=-3/16(m_{\Upsilon}-m_{\eta_b})=-11.6$ MeV and $B_{bb}\approx B^m_{bb}=-3/16[(3m_{\Upsilon _b}+m_{\eta _b})-8m^m_b]=422.0$ MeV, neglecting the small differences of constituent quark masses between baryons and mesons. The meson masses are taken from experiments and the heavy quark mass in mesons $m_{b}^m=5003.8$ MeV and $m_{b}^m=1663.2$ MeV (the superscript $m$/$b$ denotes in mesons/baryons) are from~Ref.~\cite{Karliner:2014gca} within the same model. 
For $B_{cb}$ and $\alpha_{cb}$, 
since $B^{*}_c$ is not observed yet, we have to refer to the theoretical predictions. In this work, we use the value in Ref.~\cite{Karliner:2014gca}, $m_{B^{*}_c}-m_{B_c}=68$ MeV, which is determined by an interpolation method with the mass differences $m_{\Upsilon_b}-m_{\eta_b}=62.3$ MeV and $m_{J/\psi}-m_{\eta_c}=113.2$ MeV. The four heavy meson masses are from experiments and the interpolation method used are based on the same constituent quark model as in this work. Other theoretical predictions of $m_{B^{*}_c}-m_{B_c}$ are also alternative, for example, the lattice prediction yields a value of $55(3)$ MeV~\cite{Mathur:2018epb}. Using the former value gives $B_{bc}=256.2\ \text{MeV}$ and $\alpha_{cb}/(m_bm_c)=12.7$ MeV, while the latter value yields $B_{bc}=263.5\ \text{MeV}$ and $\alpha_{cb}/(m_bm_c)=10.3$ MeV.  This results in a shift of less than $10\ \text{MeV}$ in the finally obtained $T_{bc}$ mass spectrum, which does not affect the main results and conclusions of this paper. We adopt the former value $m_{B^{*}_c}-m_{B_c}=68$ MeV for two reasons: one is based on the consideration of model consistency, and the other is that according to the heavy quark spin symmetry, the spin mass splitting of $bc$ mesons $m_{B^{*}_c}-m_{B_c}$ should lie between that of $cc$ and $bb$ mesons. 

The masses of the heavy diquarks are determined to be 
$m_{[cc]^{\bar{3}}_1}=2m^b_c+(-2/3)(B_{cc}+\alpha_{cc} /(m^b_c)^2)=3300.8$ MeV, $m_{[bb]^{\bar{3}}_1}=2m^b_b+(-2/3)(B_{bb}+\alpha_{bb} /(m^b_b)^2)=9821.0$ MeV, $m_{[cb]^{\bar{3}}_0}=m^b_c+m^b_b+(-2/3)(B_{cb}-3\alpha_{cb} /(m^b_b\cdot m^b_c))=6567.0$ MeV and $m_{[cb]^{\bar{3}}_1}=m^b_c+m^b_b+(-2/3)(B_{cb}+\alpha_{cb} /(m^b_b\cdot m^b_c))=6600.9$ MeV, respectively.
After determining the mass of the heavy diquark $m_{[QQ']^{C}_S}$, we need the binding and spin hyperfine coupling constant between ${[QQ']^{C}_S}$ and the light quark to calculate the masses of DHTs with Eq.~(\ref{MDHT}).
In Refs.~\cite{Karliner:2008sv,Karliner:2014gca,Wu:2022gie}, the studies indicate that there are no bindings between light quark pair ($qq'$) and heavy-unflavored quark pair $Qn$. 
Since the color and spin operators have been explicitly introduced in Eq.~(\ref{MDHT}), the binding between heavy diquark and light quark should be independent of both color and spin, denoted as $B_{[QQ']q}$. So we have $B_{[QQ']  q}=B_{Qq}+B_{Q'q}$, resulting in $B_{[QQ']n}=0$, $B_{[cc]s}=2B_{cs}=106.8$ MeV, $B_{[bb]s}=2B_{bs}=125.2$ MeV, and $B_{[cb]s}=B_{cs}+B_{bs}=116.0$ MeV.
For the spin hyperfine coupling constant
$\alpha_{ij}$, the constant is the same for the light-light quarks and heavy-light quarks but is flavor dependent for the heavy-heavy quarks. 
For $\alpha_{[QQ']^C_Sq}$, since the heavy diquark is regarded as a heavy quark $X$ with HADS, $\alpha_{[QQ']^C_Sq}$ should be $\alpha_{Xq}$, which is proved to be independent on the mass of heavy quark $X$ and share the same value as $\alpha_{qq'}$ (abbreviated as $\alpha$)~. For detailed discussions, refer to Refs.~\cite{Karliner:2008sv,Karliner:2014gca,Wu:2022gie}.

It should be noted that the hyperfine splitting constant $\alpha_{ij}$ is model-dependent~\cite{Luo:2017eub,Zhang:2025wmr,Wu:2022gie}, while the hyperfine splitting $\alpha_{ij}/{m_i m_j}$, combined with the spin factor,  reflects the mass splitting of different spin configurations of DHTs. According to HADS, the spin splitting should approach zero in the heavy quark limit. This is reflected in Table~I, where the spin splitting decreases as the heavy quark mass increases ($|\alpha_{cc}/m_c^2|>|\alpha_{cb}/m_cm_b| >|\alpha_{bb}/m_b^2|$). The impact of HADS symmetry breaking is also worth discussing. In this work, HADS only affects the hyperfine splitting between the heavy diquark and light quark. We therefore define $\chi_{\mathrm{HADS}}=\left|\mathcal{A}_{\mathrm{HADS}} (F_{[QQ']^{C}_{S}}\cdot F_q) ({\sigma}_{[QQ']^{C}_{S}}\cdot {\sigma}_q){\alpha_{[QQ']^{C}_{S}q}}/{m_q m_{[QQ']^{C}_{S}}}\right|$, where $\mathcal{A}_{\mathrm{HADS}}$ is defined as the degree of HADS breaking. Considering 25\% breaking of HADS~\cite{Hu:2005gf,Wu:2022gie}, the uncertainties of HADS $\chi_{\mathrm{HADS}}$ in this work are estimated to be at the level of 11.3 MeV, 5.7 MeV and 3.8 MeV for the doubly charmed, bottom-charmed and doubly bottomed tetraquarks, respectively.

The masses, bindings and spin hyperfine splittings needed for calculating the masses of DHTs are listed in Table~\ref{Paras:DHT}.  
It should be noted that no new parameters are introduced here. All parameters used to calculate the masses of DHTs are derived from the 13 independent parameters already listed in Table~\ref{Paras: Baryon}. which are all obtained by fitting experimental data of the observed hadron mass spectra. This well ensures the correlation between the theoretical predictions of DHT masses and the experimental data of baryon mass spectra, thereby enhancing the reliability of the predictions in this work.

\begin{table*}[t]
	\caption{Doubly heavy tetraquark ground states aroused from heavy anti-baryons $\bar{\Lambda}_c$($\bar{\Lambda}_b$) and $\bar{\Sigma}_c$($\bar{\Sigma}_b$) with HADS. }
    \label{table1}
	\centering
		\begin{tabular}{cccc}
			\hline\hline
	State&Configuration&$I(J^P)$&Mass formula \\
	\hline
    $T_{cc}(3876)$&$[[\bar{n}\bar{n}]^3_0[cc]^{\bar{3}}_1]^0_1$&$0(1^+)$&$2m_n^b+m_{[cc]^{\bar{3}}_1}+2\alpha /(m_n^b)^2$\\
    $T_{cc}(4035)$&$[[\bar{n}\bar{n}]^3_1[cc]^{\bar{3}}_1]^1_0$&$1(0^+)$&$2m_n^b+m_{[cc]^{\bar{3}}_1}-\frac{2}{3}[\alpha /(m_n^b)^2-8\alpha_{[cc]^{\bar{3}}_1n} /(m_n^b m_{[cc]^{\bar{3}}_1})]$\\
	$T_{cc}(4058)$&$[[\bar{n}\bar{n}]^3_1[cc]^{\bar{3}}_1]^1_1$&$1(1^+)$&$2m_n^b+m_{[cc]^{\bar{3}}_1}-\frac{2}{3}[\alpha /(m_n^b)^2-4\alpha_{[cc]^{\bar{3}}_1n} /(m_n^b m_{[cc]^{\bar{3}}_1})]$\\
			$T_{cc}(4103)$&$[[\bar{n}\bar{n}]^3_1[cc]^{\bar{3}}_1]^1_2$&$1(2^+)$&$2m_n^b+m_{[cc]^{\bar{3}}_1}-\frac{2}{3}[\alpha /(m_n^b)^2+4\alpha_{[cc]^{\bar{3}}_1n} /(m_n^b m_{[cc]^{\bar{3}}_1})]$\\
			$T_{bb}(10396)$&$[[\bar{n}\bar{n}]^3_0[bb]^{\bar{3}}_1]^0_1$&$0(1^+)$&$2m_n^b+m_{[bb]^{\bar{3}}_1}+2\alpha /(m_n^b)^2$\\
			$T_{bb}(10586)$&$[[\bar{n}\bar{n}]^3_1[bb]^{\bar{3}}_1]^1_0$&$1(0^+)$&$2m_n^b+m_{[bb]^{\bar{3}}_1}-\frac{2}{3}[\alpha /(m_n^b)^2-8\alpha_{[bb]^{\bar{3}}_1n} /(m_n^b m_{[bb]^{\bar{3}}_1})]$\\
			$T_{bb}(10593)$&$[[\bar{n}\bar{n}]^3_1[bb]^{\bar{3}}_1]^1_1$&$1(1^+)$&$2m_n^b+m_{[bb]^{\bar{3}}_1}-\frac{2}{3}[\alpha /(m_n^b)^2-4\alpha_{[bb]^{\bar{3}}_1n} /(m_n^b m_{[bb]^{\bar{3}}_1})]$\\
			$T_{bb}(10608)$&$[[\bar{n}\bar{n}]^3_1[bb]^{\bar{3}}_1]^1_2$&$1(2^+)$&$2m_n^b+m_{[bb]^{\bar{3}}_1}-\frac{2}{3}[\alpha /(m_n^b)^2+4\alpha_{[bb]^{\bar{3}}_1n} /(m_n^b m_{[bb]^{\bar{3}}_1})]$\\
			$T_{cb}(7142)$&$[[\bar{n}\bar{n}]^3_0[cb]^{\bar{3}}_0]^0_0$&$0(0^+)$&$2m^b_n+m_{[cb]^{\bar{3}}_0}+2\alpha /(m^b_n)^2$\\
			$T_{cb}(7176)$&$[[\bar{n}\bar{n}]^3_0[cb]^{\bar{3}}_1]^0_1$&$0(1^+)$&$2m^b_n+m_{[cb]^{\bar{3}}_1}+2\alpha /(m^b_n)^2$\\
			$T_{cb}(7347)$&$[[\bar{n}\bar{n}]^3_1[cb]^{\bar{3}}_0]^1_1$&$1(1^+)$&$2m^b_n+m_{[cb]^{\bar{3}}_0}-(2/3)[\alpha /(m^b_n)^2]$\\
			$T_{cb}(7358)$&$[[\bar{n}\bar{n}]^3_1[cb]^{\bar{3}}_1]^1_0$&$1(0^+)$&$2m^b_n+m_{[cb]^{\bar{3}}_1}-(2/3)[\alpha /(m^b_n)^2-8\alpha /(m^b_nm_{[cb]^{\bar{3}}_1})]$\\
			$T_{cb}(7369)$&$[[\bar{n}\bar{n}]^3_1[cb]^{\bar{3}}_1]^1_1$&$1(1^+)$&$2m^b_n+m_{[cb]^{\bar{3}}_1}-(2/3)[\alpha /(m^b_n)^2-4\alpha /(m^b_nm_{[cb]^{\bar{3}}_1})]$\\
			$T_{cb}(7392)$&$[[\bar{n}\bar{n}]^3_1[cb]^{\bar{3}}_1]^1_2$&$1(2^+)$&$2m^b_n+m_{[cb]^{\bar{3}}_1}-(2/3)[\alpha /(m^b_n)^2+4\alpha /(m^b_nm_{[cb]^{\bar{3}}_1})]$\\
			
			\hline\hline
		\end{tabular}
\end{table*}

\begin{table*}[t]
	\caption{Doubly heavy tetraquark ground states aroused from heavy anti-baryons $\bar{\Omega}_c$($\bar{\Omega}_b$) with HADS.}
    \label{table2}
	\centering
		\begin{tabular}{cccc}
			\hline\hline
   State&Configuration&$J^P$&Mass formula\\
			\hline
			$T_{cc\bar{s}\bar{s}}(4224)$&$[[\bar{s}\bar{s}]^{3}_1[cc]^{\bar{3}}_1]_0$&$0^+$&$2m^b_s+m_{[cc]^{\bar{3}}_1}-(2/3)[2B_{[cc]^{\bar{3}}_1s}+\alpha /(m^b_s)^2-8\alpha /(m^b_sm_{[cc]^{\bar{3}}_1})]$\\
			$T_{cc\bar{s}\bar{s}}(4239)$&$[[\bar{s}\bar{s}]^{3}_1[cc]^{\bar{3}}_1]_1$&$1^+$&$2m^b_s+m_{[cc]^{\bar{3}}_1}-(2/3)[2B_{[cc]^{\bar{3}}_1s}+\alpha /(m^b_s)^2-4\alpha /(m^b_sm_{[cc]^{\bar{3}}_1})]$\\
			$T_{cc\bar{s}\bar{s}}(4270)$&$[[\bar{s}\bar{s}]^{3}_1[cc]^{\bar{3}}_1]_2$&$2^+$&$2m^b_s+m_{[cc]^{\bar{3}}_1}-(2/3)[2B_{[cc]^{\bar{3}}_1s}+\alpha /(m^b_s)^2+4\alpha /(m^b_sm_{[cc]^{\bar{3}}_1})]$\\
			$T_{bb\bar{s}\bar{s}}(10740)$&$[[\bar{s}\bar{s}]^{3}_1[bb]^{\bar{3}}_1]_0$&$0^+$&$2m^b_s+m_{[bb]^{\bar{3}}_1}-(2/3)[2B_{[bb]^{\bar{3}}_1s}+\alpha /(m^b_s)^2-8\alpha /(m^b_sm_{[bb]^{\bar{3}}_1})]$\\
			$T_{bb\bar{s}\bar{s}}(10745)$&$[[\bar{s}\bar{s}]^{3}_1[bb]^{\bar{3}}_1]_1$&$1^+$&$2m^b_s+m_{[bb]^{\bar{3}}_1}-(2/3)[2B_{[bb]^{\bar{3}}_1s}+\alpha /(m^b_s)^2-4\alpha /(m^b_sm_{[bb]^{\bar{3}}_1})]$\\
			$T_{bb\bar{s}\bar{s}}(10755)$&$[[\bar{s}\bar{s}]^{3}_1[[bb]^{\bar{3}}_1]_2$&$2^+$&$2m^b_s+m_{[bb]^{\bar{3}}_1}-(2/3)[2B_{[bb]^{\bar{3}}_1s}+\alpha /(m^b_s)^2+4\alpha /(m^b_sm_{[bb]^{\bar{3}}_1})]$\\
			$T_{cb\bar{s}\bar{s}}(7508)$&$[[\bar{s}\bar{s}]^{3}_1[cb]^{\bar{3}}_0]_1$&$1^+$&$2m^b_s+m_{[cb]^{\bar{3}}_0}-(2/3)[2B_{[cb]^{\bar{3}}_0s}+\alpha /(m^b_s)^2]$\\
			$T_{cb\bar{s}\bar{s}}(7527)$&$[[\bar{s}\bar{s}]^{3}_1[cb]^{\bar{3}}_1]_0$&$0^+$&$2m^b_s+m_{[cb]^{\bar{3}}_1}-(2/3)[2B_{[cb]^{\bar{3}}_1s}+\alpha /(m^b_s)^2-8\alpha /(m^b_sm_{[cb]^{\bar{3}}_1})]$\\
			$T_{cb\bar{s}\bar{s}}(7535)$&$[[\bar{s}\bar{s}]^{3}_1[cb]^{\bar{3}}_1]_1$&$1^+$&$2m^b_s+m_{[cb]^{\bar{3}}_1}-(2/3)[2B_{[cb]^{\bar{3}}_1s}+\alpha /(m^b_s)^2-4\alpha /(m^b_sm_{[cb]^{\bar{3}}_1})]$\\
			$T_{cb\bar{s}\bar{s}}(7550)$&$[[\bar{s}\bar{s}]^{3}_1[cb]^{\bar{3}}_1]_2$&$2^+$&$2m^b_s+m_{[cb]^{\bar{3}}_1}-(2/3)[2B_{[cb]^{\bar{3}}_1s}+\alpha /(m^b_s)^2+4\alpha /(m^b_sm_{[cb]^{\bar{3}}_1})]$\\
			\hline\hline
			\end{tabular}
	\end{table*}

\begin{table*}[t]
\caption{Doubly heavy tetraquark ground states aroused from heavy anti-baryons $\bar{\Xi}_c$($\bar{\Xi}_b$) with HADS.}
\label{table3}
	\centering
		\begin{tabular}{cccc}
			\hline\hline
	State&Configuration&$I(J^P)$&Mass formula \\
			\hline
			$T_{cc\bar{s}}(4026)$&$[[\bar{n}\bar{s}]^{3}_0[cc]^{\bar{3}}_1]^{\frac{1}{2}}_1$&$\frac{1}{2}(1^+)$&$m^b_n+m^b_s+m_{[cc]^{\bar{3}}_1}-(2/3)[B_{[cc]^{\bar{3}}_1s}-3\alpha /(m^b_nm^b_s)]$\\
			$T_{cc\bar{s}}(4127)$&$[[\bar{n}\bar{s}]^{3}_1[cc]^{\bar{3}}_1]^{\frac{1}{2}}_0$&$\frac{1}{2}(0^+)$&$m^b_n+m^b_s+m_{[cc]^{\bar{3}}_1}-(2/3)[B_{[cc]^{\bar{3}}_1s}+\alpha /(m^b_nm^b_s)-4\alpha /(m^b_nm_{[cc]^{\bar{3}}_1})-4\alpha /(m^b_sm_{[cc]^{\bar{3}}_1})]$\\
			$T_{cc\bar{s}}(4146)$&$[[\bar{n}\bar{s}]^{3}_1[cc]^{\bar{3}}_1]^{\frac{1}{2}}_1$&$\frac{1}{2}(1^+)$&$m^b_n+m^b_s+m_{[cc]^{\bar{3}}_1}-(2/3)[B_{[cc]^{\bar{3}}_1s}+\alpha /(m^b_nm^b_s)-2\alpha /(m^b_nm_{[cc]^{\bar{3}}_1})-2\alpha /(m^b_sm_{[cc]^{\bar{3}}_1})]$\\
			$T_{cc\bar{s}}(4184)$&$[[\bar{n}\bar{s}]^{3}_1[cc]^{\bar{3}}_1]^{\frac{1}{2}}_2$&$\frac{1}{2}(2^+)$&$m^b_n+m^b_s+m_{[cc]^{\bar{3}}_1}-(2/3)[B_{[cc]^{\bar{3}}_1s}+\alpha /(m^b_nm^b_s)+2\alpha /(m^b_nm_{[cc]^{\bar{3}}_1})+2\alpha /(m^b_sm_{[cc]^{\bar{3}}_1})]$\\
			$T_{bb\bar{s}}(10534)$&$[[\bar{n}\bar{s}]^{3}_0[bb]^{\bar{3}}_1]^{\frac{1}{2}}_1$&$\frac{1}{2}(1^+)$&$m^b_n+m^b_s+m_{[bb]^{\bar{3}}_1}-(2/3)[B_{[bb]^{\bar{3}}_1s}-3\alpha /(m^b_nm^b_s)]$\\
			$T_{bb\bar{s}}(10660)$&$[[\bar{n}\bar{s}]^{3}_1[bb]^{\bar{3}}_1]^{\frac{1}{2}}_0$&$\frac{1}{2}(0^+)$&$m^b_n+m^b_s+m_{[bb]^{\bar{3}}_1}-(2/3)[B_{[bb]^{\bar{3}}_1s}+\alpha /(m^b_nm^b_s)-4\alpha /(m^b_nm_{[bb]^{\bar{3}}_1})-4\alpha /(m^b_sm_{[bb]^{\bar{3}}_1})]$\\
			$T_{bb\bar{s}}(10666)$&$[[\bar{n}\bar{s}]^{3}_1[bb]^{\bar{3}}_1]^{\frac{1}{2}}_1$&$\frac{1}{2}(1^+)$&$m^b_n+m^b_s+m_{[bb]^{\bar{3}}_1}-(2/3)[B_{[bb]^{\bar{3}}_1s}+\alpha /(m^b_nm^b_s)-2\alpha /(m^b_nm_{[bb]^{\bar{3}}_1})-2\alpha /(m^b_sm_{[bb]^{\bar{3}}_1})]$\\
			$T_{bb\bar{s}}(10679)$&$[[\bar{n}\bar{s}]^{3}_1[bb]^{\bar{3}}_1]^{\frac{1}{2}}_2$&$\frac{1}{2}(2^+)$&$m^b_n+m^b_s+m_{[bb]^{\bar{3}}_1}-(2/3)[B_{[bb]^{\bar{3}}_1s}+\alpha /(m^b_nm^b_s)+2\alpha /(m^b_nm_{[bb]^{\bar{3}}_1})+2\alpha /(m^b_sm_{[bb]^{\bar{3}}_1})]$\\
			$T_{cb\bar{s}}(7286)$&$[[\bar{n}\bar{s}]^{3}_0[cb]^{\bar{3}}_0]^{\frac{1}{2}}_0$&$\frac{1}{2}(0^+)$&$m^b_n+m^b_s+m_{[cb]^{\bar{3}}_0}-(2/3)[B_{[cb]^{\bar{3}}_0s}-3\alpha /(m^b_nm^b_s)]$\\
			$T_{cb\bar{s}}(7320)$&$[[\bar{n}\bar{s}]^{3}_0[cb]^{\bar{3}}_1]^{\frac{1}{2}}_1$&$\frac{1}{2}(1^+)$&$m^b_n+m^b_s+m_{[cb]^{\bar{3}}_1}-(2/3)[B_{[cb]^{\bar{3}}_1s}-3\alpha /(m^b_nm^b_s)]$\\
			$T_{cb\bar{s}}(7425)$&$[[\bar{n}\bar{s}]^{3}_1[cb]^{\bar{3}}_0]^{\frac{1}{2}}_1$&$\frac{1}{2}(1^+)$&$m^b_n+m^b_s+m_{[cb]^{\bar{3}}_0}-(2/3)[B_{[cb]^{\bar{3}}_0s}+\alpha /(m^b_nm^b_s)]$\\
			$T_{cb\bar{s}}(7440)$&$[[\bar{n}\bar{s}]^{3}_1[cb]^{\bar{3}}_1]^{\frac{1}{2}}_0$&$\frac{1}{2}(0^+)$&$m^b_n+m^b_s+m_{[cb]^{\bar{3}}_1}-(2/3)[B_{[cb]^{\bar{3}}_1s}+\alpha /(m^b_nm^b_s)-4\alpha /(m^b_nm_{[cb]^{\bar{3}}_1})-4\alpha /(m^b_sm_{[cb]^{\bar{3}}_1})]$\\
			$T_{cb\bar{s}}(7449)$&$[[\bar{n}\bar{s}]^{3}_1[cb]^{\bar{3}}_1]^{\frac{1}{2}}_1$&$\frac{1}{2}(1^+)$&$m^b_n+m^b_s+m_{[cb]^{\bar{3}}_1}-(2/3)[B_{[cb]^{\bar{3}}_1s}+\alpha /(m^b_nm^b_s)-2\alpha /(m^b_nm^b_{[cb]^{\bar{3}}_1})-2\alpha /(m^b_sm_{[cb]^{\bar{3}}_1})]$\\
			$T_{cb\bar{s}}(7468)$&$[[\bar{n}\bar{s}]^{3}_1[cb]^{\bar{3}}_1]^{\frac{1}{2}}_2$&$\frac{1}{2}(2^+)$&$m^b_n+m^b_s+m_{[cb]^{\bar{3}}_1}-(2/3)[B_{[cb]^{\bar{3}}_1s}+\alpha /(m^b_nm^b_s)+2\alpha /(m^b_nm_{[cb]^{\bar{3}}_1})+2\alpha /(m^b_sm_{[cb]^{\bar{3}}_1})]$\\

			\hline\hline
			\end{tabular}
	\end{table*}

\section{Results and discussions}
\label{Results}

In this section, we use the mass formula introduced in the last section with the parameters listed in Table~\ref{Paras:DHT} to calculate the mass spectra of DHTs. 
Based on different combinations of quark flavors and quantum numbers, we predict and calculate the masses of 38 DHT states linked to heavy baryons through HADS. This includes: 
\begin{itemize}
    \item {14 DHT partners} of $\Lambda_Q$ and $\Sigma_Q$ baryons: 
    \begin{itemize}
        \item 6 partners for isospin-0 $\Lambda_Q$ baryons
        \item 8 partners for isospin-1 $\Sigma_Q$ baryons
    \end{itemize}
    \item {14 DHT partners} of $\Xi_Q$ baryons
    \item {10 DHT partners} of $\Omega_Q$ baryons
\end{itemize}

 The masses, quantum numbers, configurations and mass formulas of the 38 predicted DHTs are systematically categorized and presented across Tables~\ref{table1}, \ref{table2}, and~\ref{table3}.  It should be noted that the states listed in the tables are calculated one-to-one according to configurations using the mass formula. Actually, since certain configurations share the same quantum numbers and quark components, mixing can occur between them. We can compute the correlation matrix of states with different configurations to determine the eigenmasses and eigenvectors of the mixed eigenstates, thereby obtaining the mass spectrum of DHT states considering mixing effects. 
The diagonal elements of the correlation matrix correspond to the masses of explicit configurations given in Tables~\ref{table1}, \ref{table2}, and~\ref{table3}, while the off-diagonal elements represent the couplings between different configurations under the mass Hamiltonian (i.e., Eq.(\ref{MDHT})). This coupling originates from the different projections of color operators and spin operators onto different configurations.
Considering that we adopt HADS in this work, where the heavy diquark is treated as an integral anti-heavy quark, the configurations $[QQ']_1^3$ and $[QQ']_0^3$ are regarded as distinct ``particles" and cannot mix. Consequently, the mixing of DHT states is simplified to a form similar to the mixing of three-quark heavy baryon configurations—this is a key distinction between the present work and other studies. By comparing the mass spectrum of mixed states with and without the HADS approximation, we can assess the impact of introducing HADS on the DHT mass spectrum.

In Table~\ref{Mix1}, we present the mass correlation matrices of different configurations under the same quantum numbers considering HADS, along with the final eigenmasses and eigenvectors. The results indicate that mixing primarily occurs for the quantum number $1/2(1^+)$: $T_{cc}$, $T_{bb}$, and $T_{bc}$ each have two configurations that undergo mixing, with mass differences of approximately several MeV to over ten MeV compared to the unmixed cases.
In the alternative scenario where we do not consider the HADS approximation and still treat DHT as a four-body system, the projection values of color operators and spin operators can be found in Tables II and III of Ref.~\cite{Lu:2020rog}. The results for this case are listed in Table~\ref{Mix2}. Compared to the HADS-included case, $T_{cb}$ has seven additional mixed configurations, mainly due to the allowable mixing between $[cb]_1^{\bar{3}}$ and $[cb]_0^{\bar{3}}$ without HADS. A comparison of the results in Tables~\ref{Mix1} and ~\ref{Mix2} shows that the introduction of HADS has an impact of approximately ten-odd MeV on the masses of mixed DHT states, consistent with the aforementioned analysis.

\begin{table*}[t]
	\caption{The eigenmasses and eigenvectors of the DHT states with quantum numbers and mixing configurations with HADS.}
    \label{Mix1}
	\centering
		\begin{tabular}{ccccc}
         \hline\hline
$I(J^p)$&Configuration&$\langle H\rangle$ (MeV)&Mass (MeV)&Eigenvector\\
            \hline
$\frac{1}{2}(1^+)$&$[[\bar{n}\bar{s}]^3_0[cc]^{\bar{3}}_1]^{\frac{1}{2}}_1$ &\multirow{2}{*}{$\begin{pmatrix}
    4026&40\\40&4146
\end{pmatrix}$} &\multirow{2}{*}{$\begin{bmatrix}
    4014\\4158
\end{bmatrix}$}&\multirow{2}{*}{$\begin{bmatrix}
    (-0.957,0.290))\\(0.290,0.957)
\end{bmatrix}$}\\
&$[[\bar{n}\bar{s}]^3_1[cc]^{\bar{3}}_1]^{\frac{1}{2}}_1$&&\\
$\frac{1}{2}(1^+)$&$[[\bar{n}\bar{s}]^3_0[bb]^{\bar{3}}_1]^{\frac{1}{2}}_1$ &\multirow{2}{*}{$\begin{pmatrix}
    10534&13\\13&10666
\end{pmatrix}$} &\multirow{2}{*}{$\begin{bmatrix}
    10533\\10667
\end{bmatrix}$}&\multirow{2}{*}{$\begin{bmatrix}
    (-0.995,0.097))\\(0.097,0.995)
\end{bmatrix}$}\\
&$[[\bar{n}\bar{s}]^3_1[bb]^{\bar{3}}_1]^{\frac{1}{2}}_1$&&\\
$\frac{1}{2}(1^+)$&$[[\bar{n}\bar{s}]^3_0[cb]^{\bar{3}}_1]^{\frac{1}{2}}_1$ &\multirow{2}{*}{$\begin{pmatrix}
    7320&20\\20&7449
\end{pmatrix}$} &\multirow{2}{*}{$\begin{bmatrix}
    7317\\7452
\end{bmatrix}$}&\multirow{2}{*}{$\begin{bmatrix}
    (-0.989,0.150))\\(0.150,0.989)
\end{bmatrix}$}\\
&$[[\bar{n}\bar{s}]^3_1[cb]^{\bar{3}}_1]^{\frac{1}{2}}_1$&&\\
			\hline\hline
			\end{tabular}
	\end{table*}

\begin{table*}[]
	\caption{The eigenmasses and eigenvectors of the DHT states with quantum numbers and mixing configurations without HADS.}
    \label{Mix2}
	\centering
		\begin{tabular}{ccccc}
         \hline\hline
$I(J^p)$&Configuration&$\langle H\rangle$ (MeV)&Mass (MeV)&Eigenvector\\
            \hline
$\frac{1}{2}(1^+)$&$[[\bar{n}\bar{s}]^3_0[cc]^{\bar{3}}_1]^{\frac{1}{2}}_1$ &\multirow{2}{*}{$\begin{pmatrix}
    4026&-10\\-10&4128
\end{pmatrix}$} &\multirow{2}{*}{$\begin{bmatrix}
    4025\\4129
\end{bmatrix}$}&\multirow{2}{*}{$\begin{bmatrix}
    (0.995,0.097))\\(-0.097,0.995)
\end{bmatrix}$}\\
&$[[\bar{n}\bar{s}]^3_1[cc]^{\bar{3}}_1]^{\frac{1}{2}}_1$&&\\
$\frac{1}{2}(1^+)$&$[[\bar{n}\bar{s}]^3_0[bb]^{\bar{3}}_1]^{\frac{1}{2}}_1$ &\multirow{2}{*}{$\begin{pmatrix}
    10534&-3\\-3&10660
\end{pmatrix}$} &\multirow{2}{*}{$\begin{bmatrix}
    10534\\10660
\end{bmatrix}$}&\multirow{2}{*}{$\begin{bmatrix}
    (0.999,0.023))\\(-0.023,0.999)
\end{bmatrix}$}\\
&$[[\bar{n}\bar{s}]^3_0[bb]^{\bar{3}}_1]^{\frac{1}{2}}_1$&&\\
$1(1^+)$&$[[\bar{n}\bar{n}]^3_1[cb]^{\bar{3}}_0]^{1}_1$ 
&\multirow{2}{*}{$\begin{pmatrix}
    7347&20\\20&7351
\end{pmatrix}$} &\multirow{2}{*}{$\begin{bmatrix}
    7329\\7369
\end{bmatrix}$}&\multirow{2}{*}{$\begin{bmatrix}
    (-0.741,0.671))\\(0.671,0.741)
\end{bmatrix}$}\\
&$[[\bar{n}\bar{n}]^3_1[cb]^{\bar{3}}_1]^{1}_1$&&\\
$\frac{1}{2}(0^+)$&$[[\bar{n}\bar{s}]^3_0[cb]^{\bar{3}}_0]^{\frac{1}{2}}_0$ &\multirow{2}{*}{$\begin{pmatrix}
    7286&-4\\-4&7410
\end{pmatrix}$} &\multirow{2}{*}{$\begin{bmatrix}
    7286\\7410
\end{bmatrix}$}&\multirow{2}{*}{$\begin{bmatrix}
    (0.999,0.032))\\(-0.032,0.999)
\end{bmatrix}$}\\
&$[[\bar{n}\bar{s}]^3_1[cb]^{\bar{3}}_1]^{\frac{1}{2}}_0$&&\\
$\frac{1}{2}(1^+)$&$[[\bar{n}\bar{s}]^3_0[cb]^{\bar{3}}_1]^{\frac{1}{2}}_1$ &\multirow{2}{*}{$\begin{pmatrix}
    7320&2&-7\\2&7425&17\\-7&17&7434
\end{pmatrix}$} &\multirow{3}{*}{$\begin{bmatrix}
    7319\\7412\\7447
\end{bmatrix}$}&\multirow{3}{*}{$\begin{bmatrix}
    (0.997,-0.029,0.065))\\(-0.063,-0.796,0.603)\\(-0.034,0.605,0.795)
\end{bmatrix}$}\\
&$[[\bar{n}\bar{s}]^3_1[cb]^{\bar{3}}_0]^{\frac{1}{2}}_1$&&\\
&$[[\bar{n}\bar{s}]^3_1[cb]^{\bar{3}}_1]^{\frac{1}{2}}_1$&&\\
$(1^+)$&$[[\bar{s}\bar{s}]^3_1[cb]^{\bar{3}}_0]_1$ &\multirow{2}{*}{$\begin{pmatrix}
    7508&14\\14&7522
\end{pmatrix}$} &\multirow{2}{*}{$\begin{bmatrix}
    7499\\7531
\end{bmatrix}$}&\multirow{2}{*}{$\begin{bmatrix}
    (-0.851,0.526))\\(0.526,0.851)
\end{bmatrix}$}\\
&$[[\bar{s}\bar{s}]^3_1[cb]^{\bar{3}}_1]_1$&&\\
			\hline\hline
			\end{tabular}
	\end{table*}

The masses and quantum numbers of the 38 predicted DHTs with mixing effect are categorized and presented across Tables~\ref{TQQ'qq}, \ref{TQQ'sq}, and~\ref{TQQ'ss}. These encompass three distinct groups: 14 states linked to $\bar{\Lambda}_c$/$\bar{\Lambda}_b$ and $\bar{\Sigma}_c$/$\bar{\Sigma}_b$ baryons, 14 states associated with $\bar{\Xi}_c$/$\bar{\Xi}_b$ baryons, and 10 states corresponding to $\bar{\Omega}_c$/$\bar{\Omega}_b$ baryons. 
Each table provides comparative mass predictions from multiple theoretical approaches, revealing an overall consistency in spectral hierarchies—particularly the mass ordering $T_{QQ'} < T_{QQ'\bar{s} }< T_{QQ'\bar{s}\bar{s}}$—while highlighting method-dependent variations in absolute values.  
Notably, our predictions derive from a framework exploiting HADS to establish rigorous connections between tetraquark spectra and experimentally measured heavy baryon spectrum. This approach contrasts with complementary methodologies: relativistic quark models employing diquark-antidiquark configurations~\cite{Ebert:2007rn} or variational methods~\cite{Lu:2020rog}; effective interaction models including color-magnetic interactions~\cite{Luo:2017eub} and extended chromomagnetic model~\cite{Weng:2021hje}; and non-perturbative techniques such as diffusion Monte Carlo~\cite{Mutuk:2023oyz} and MIT bag model calculations~\cite{Zhang:2021yul}. 

The tabulated comparisons demonstrate that our HADS-constrained predictions show closer agreement with near-threshold states like $T_{cc}(3876)$, while providing systematic coverage of strange quark effects across all configurations. Unlike molecular models that struggle with hyperfine splittings, our approach delivers precise mass differences between spin partners—a critical discriminator for exotic state identification. These comprehensive spectral predictions establish reliable benchmarks for experimental searches at LHCb and Belle II, while underscoring HADS as a fundamental symmetry bridging conventional baryon and exotic tetraquark spectroscopy.

\begin{table*}[htbp]
\centering
\caption{Spectrum of DHTs aroused from heavy anti-baryons $\bar{\Lambda}_c$($\bar{\Lambda}_b$) and $\bar{\Sigma}_c$($\bar{\Sigma}_b$) with HADS (values in brackets are without HADS) in this work and comparisons with different models or methods. DMC: diffusion Monte Carlo method. CMI: color-magnetic interaction model; RQM(VM): relativized quark model within the variational method; RQM(DA): relativized quark model within diquark-antidiquark picture; MIT: MIT bag model; ECM: extended chromomagnetic model.}
\begin{tabular}{cccccccccc}
\hline\hline
$I(J^P)$ & Configuration & This work & DMC~\cite{Mutuk:2023oyz} & CMI~\cite{Luo:2017eub} & MIT~\cite{Zhang:2021yul} & RQM(VM)~\cite{Lu:2020rog} & RQM(DA)~\cite{Ebert:2007rn} & ECM I~\cite{Weng:2021hje} & ECM II~\cite{Weng:2021hje} \\
\hline
$0(1^+)$ & $[[\bar{n}\bar{n}]^3_0[cc]^{\bar{3}}_1]^0_1$ &  $3876$ & $3892$ & $4007$ & $3925$ & $4041$ & $3935$ & $3749.8$ & $3868.7$ \\
$1(0^+)$ & $[[\bar{n}\bar{n}]^3_1[cc]^{\bar{3}}_1]^1_0$ &  $4035$ & $4062$ & $4078$ & $4032$ & $4159$ & $4056$ & $3833.2$ & $3969.2$ \\
$1(1^+)$ & $[[\bar{n}\bar{n}]^3_1[cc]^{\bar{3}}_1]^1_1$ &  $4058$ & $4104$ & $4021$ & $4117$ & $4268$ & $4079$ & $3946.4$ & $4053.2$ \\
$1(2^+)$ & $[[\bar{n}\bar{n}]^3_1[cc]^{\bar{3}}_1]^1_2$ & $4103$ & $4207$ & $4271$ & $4179$ & $4318$ & $4118$ & $4017.1$ & $4123.8$ \\
$0(1^+)$ & $[[\bar{n}\bar{n}]^3_0[bb]^{\bar{3}}_1]^0_1$ & $10396$ & $10,338$ & $10686$ & $10654$ & $10550$ & $10502$ & $10291.6$ & $10390.9$ \\
$1(0^+)$ & $[[\bar{n}\bar{n}]^3_1[bb]^{\bar{3}}_1]^1_0$ &  $10586$ & $10,624$ & $10841$ & $10834$ & $10765$ & $10648$ & $10468.8$ & $10569.3$ \\
$1(1^+)$ & $[[\bar{n}\bar{n}]^3_1[bb]^{\bar{3}}_1]^1_1$ &  $10593$ & $10,680$ & $10875$ & $10854$ & $10779$ & $10657$ & $10485.3$ & $10584.2$ \\
$1(2^+)$ & $[[\bar{n}\bar{n}]^3_1[bb]^{\bar{3}}_1]^1_2$ &  $10608$ & $10702$ & $10897$ & $10878$ & $10799$ & $10673$ & $10507.9$ & $10606.8$ \\
$0(0^+)$ & $[[\bar{n}\bar{n}]^3_0[cb]^{\bar{3}}_0]^0_0$ &  $7142$ & $7108$ & $7256$ & $7260$ & $7297$ & $7239$ & $7003.4$ & $7124.6$ \\
$0(1^+)$ & $[[\bar{n}\bar{n}]^3_0[cb]^{\bar{3}}_1]^0_1$ &  $7176$ & $7084$ & $7321$ & $7288$ & $7325$ & $7246$ & $7046.2$ & $7158.0$ \\
$1(0^+)$ &$[[\bar{n}\bar{n}]^3_1[cb]^{\bar{3}}_1]^1_0$ &  $735$ & $7396$ & $7457$ & $7438$ & $7519$ & $7383$ & $7189.5$ & $7305.6$ \\
$1(1^+)$ & $[[\bar{n}\bar{n}]^3_1[cb]^{\bar{3}}_0]^1_1$ &  $7347(7329)$ & $7510$ & $7473$ & $7465$ & $7537$ & $7396$ & $7211.0$ & $7322.5$ \\
&$[[\bar{n}\bar{n}]^3_1[cb]^{\bar{3}}_1]^1_1$ & $7369(7369)$ & $7542$ & $7548$ & $7509$ & $7561$ & $7403$ & $7211.0$ & $7322.5$ \\
$1(2^+)$ &$[[\bar{n}\bar{n}]^3_1[cb]^{\bar{3}}_1]^1_2$ &  $7392$ & $7568$ & $7582$ & $7531$ & $7586$ & $7422$ & $7293.2$ & $7396.0$ \\
\hline\hline
\end{tabular}
\label{TQQ'qq}
\end{table*}

\begin{table*}[htbp]
\centering
\caption{Spectrum of DHTs aroused from heavy anti-baryons $\bar{\Xi}_c$($\bar{\Xi}_b$) with HADS (values in brackets are without HADS) in this work and comparisons with different models or methods. DMC: diffusion Monte Carlo method. CMI: color-magnetic interaction model; RQM(VM): relativized quark model within the variational method; RQM(DA): relativized quark model within diquark-antidiquark picture; MIT: MIT bag model; ECM: extended chromomagnetic model.}
\begin{tabular}{cccccccccc}
\hline\hline
$I(J^P)$ & Configuration & This work & DMC~\cite{Mutuk:2023oyz} & CMI~\cite{Luo:2017eub} & MIT~\cite{Zhang:2021yul} & RQM(VM)~\cite{Lu:2020rog} & RQM(DA)~\cite{Ebert:2007rn} & ECM I~\cite{Weng:2021hje} & ECM II~\cite{Weng:2021hje} \\
\hline
$\frac{1}{2}(0^+)$ & $[[\bar{n}\bar{s}]^{3}_1[cc]^{\bar{3}}_1]^{\frac{1}{2}}_0$ & $4127$ & $4165$ & $4236$ & $4165$ & $4323$ & $4221$ & $3937.6$ & $4085.7$ \\
$\frac{1}{2}(1^+)$ & $[[\bar{n}\bar{s}]^{3}_0[cc]^{\bar{3}}_1]^{\frac{1}{2}}_1$ & $4014(4025)$ & $4056$ & $4225$ & $4091$ & $4232$ & $4143$ & $3919.0$ & $4051.5$ \\
& $[[\bar{n}\bar{s}]^{3}_1[cc]^{\bar{3}}_1]^{\frac{1}{2}}_1$ & $4158(4129)$ & $4128$ & $4363$ & $4247$ & $4394$ & $4239$ & $4073.0$ & $4180.2$ \\
$\frac{1}{2}(2^+)$ & $[[\bar{n}\bar{s}]^{3}_1[cc]^{\bar{3}}_1]^{\frac{1}{2}}_2$ & $4184$ & $4314$ & $4434$ & $4305$ & $4440$ & $4271$ & $4144.3$ & $4251.5$ \\
$\frac{1}{2}(0^+)$ & $[[\bar{n}\bar{s}]^{3}_1[bb]^{\bar{3}}_1]^{\frac{1}{2}}_0$ & $10660$ & $10721$ & $10999$ & $10955$ & $10883$ & $10802$ & $10586.4$ & $10684.1$ \\
$\frac{1}{2}(1^+)$ & $[[\bar{n}\bar{s}]^{3}_0[bb]^{\bar{3}}_1]^{\frac{1}{2}}_1$ & $10533(10534)$ & $10684$ & $10911$ & $10811$ & $10734$ & $10706$ & $10473.1$ & $10569.0$ \\
& $[[\bar{n}\bar{s}]^{3}_1[bb]^{\bar{3}}_1]^{\frac{1}{2}}_1$ & $10667(10660)$ & $10736$ & $11010$ & $10974$ & $10897$ & $10809$ & $10605.3$ & $10700.5$ \\
$\frac{1}{2}(2^+)$ & $[[\bar{n}\bar{s}]^{3}_1[bb]^{\bar{3}}_1]^{\frac{1}{2}}_2$ & $10679$ & $10747$ & $11060$ & $10997$ & $10915$ & $10823$ & $10628.7$ & $10723.9$ \\
$\frac{1}{2}(0^+)$ & $[[\bar{n}\bar{s}]^{3}_0[cb]^{\bar{3}}_0]^{\frac{1}{2}}_0$ & $7286(7286)$ & $-$ & $7461$ & $-$ & $7483$ & $7444$ & $7156.5$ & $7296.2$ \\
& $[[\bar{n}\bar{s}]^{3}_1[cb]^{\bar{3}}_1]^{\frac{1}{2}}_0$ & $7440(7410)$ & $-$ & $7615$ & $-$ & $7643$ & $7540$ & $7299.0$ & $7421.0$ \\
$\frac{1}{2}(1^+)$ & $[[\bar{n}\bar{s}]^{3}_0[cb]^{\bar{3}}_1]^{\frac{1}{2}}_1$ & $7317(7319)$ & $-$ & $7530$ & $-$ & $7514$ & $7451$ & $7212.8$ & $7336.1$ \\
& $[[\bar{n}\bar{s}]^{3}_1[cb]^{\bar{3}}_1]^{\frac{1}{2}}_1$ & $7452(7447)$ & $-$ & $7706$ & $-$ & $7682$ & $7552$ & $7330.5$ & $7487.8$ \\
& $[[\bar{n}\bar{s}]^{3}_1[cb]^{\bar{3}}_0]^{\frac{1}{2}}_1$ & $7425(7412)$ & $-$ & $7634$ & $-$ & $7659$ & $7555$ & $7323.9$ & $7440.9$ \\
$\frac{1}{2}(2^+)$ & $[[\bar{n}\bar{s}]^{3}_1[cb]^{\bar{3}}_1]^{\frac{1}{2}}_2$ & $7468$ & $-$ & $7718$ & $-$ & $7705$ & $7572$ & $7415.1$ & $7517.1$ \\
\hline\hline
\end{tabular}
\label{TQQ'sq}
\end{table*}

\begin{table*}[htbp]
\centering
\caption{Spectrum of DHTs aroused from heavy anti-baryons $\bar{\Omega}_c$($\bar{\Omega}_b$) with HADS (values in brackets are without HADS) in this work and comparisons with different models or methods. DMC: diffusion Monte Carlo method. CMI: color-magnetic interaction model; RQM(VM): relativized quark model within the variational method; RQM(DA): relativized quark model within diquark-antidiquark picture; MIT: MIT bag model; ECM: extended chromomagnetic model.}
\begin{tabular}{cccccccccc}
\hline\hline
$J^P$ & Configuration & This work & DMC~\cite{Mutuk:2023oyz} & CMI~\cite{Luo:2017eub} & MIT~\cite{Zhang:2021yul} & RQM(VM)~\cite{Lu:2020rog} & RQM(DA)~\cite{Ebert:2007rn} & ECM I~\cite{Weng:2021hje} & ECM II~\cite{Weng:2021hje} \\
\hline
$0^+$ & $[[\bar{s}\bar{s}]^{3}_1[cc]^{\bar{3}}_1]_0$ &  $4224$ & $4205$ & $4395$ & $4300$ & $4417$ & $4359$ & $4043.7$ & $4199.1$ \\
$1^+$ & $[[\bar{s}\bar{s}]^{3}_1[cc]^{\bar{3}}_1]_1$ & $4239$ & $4323$ & $4526$ & $4382$ & $4493$ & $4375$ & $4192.6$ & $4300.2$ \\
$2^+$ & $[[\bar{s}\bar{s}]^{3}_1[cc]^{\bar{3}}_1]_2$ & $4270$ & $4381$ & $4597$ & $4433$ & $4536$ & $4402$ & $4264.5$ & $4372.1$ \\
$0^+$ & $[[\bar{s}\bar{s}]^{3}_1[bb]^{\bar{3}}_1]_0$ & $10740$ & $10868$ & $11157$ & $11078$ & $10972$ & $10932$ & $10697.1$ & $10792.1$ \\
$1^+$ & $[[\bar{s}\bar{s}]^{3}_1[bb]^{\bar{3}}_1]_1$ & $10745$ & $10908$ & $11199$ & $11099$ & $10986$ & $10939$ & $10718.2$ & $10809.8$ \\
$2^+$ & $[[\bar{s}\bar{s}]^{3}_1[bb]^{\bar{3}}_1]_2$ & $10755$ & $10926$ & $11224$ & $11119$ & $11004$ & $10950$ & $10742.5$ & $10834.1$ \\
$0^+$ & $[[\bar{s}\bar{s}]^{3}_1[cb]^{\bar{3}}_1]_0$ & $7527$ & $7653$ & $7774$ & $7693$ & $7735$ & $7673$ & $7404.4$ & $7531.3$ \\
$1^+$ & $[[\bar{s}\bar{s}]^{3}_1[cb]^{\bar{3}}_0]_1$ & $7508(7499)$ & $7711$ & $7793$ & $7716$ & $7752$ & $7683$ & $7431.8$ & $7553.6$ \\
& $[[\bar{s}\bar{s}]^{3}_1[cb]^{\bar{3}}_1]_1$ & $7535(7531)$ & $7764$ & $7872$ & $7757$ & $7775$ & $7684$ & $7503.3$ & $7603.5$ \\
$2^+$ & $[[\bar{s}\bar{s}]^{3}_1[cb]^{\bar{3}}_1]_2$ & $7550$ & $7862$ & $7908$ & $7779$ & $7798$ & $7701$ & $7534.3$ & $7633.8$ \\
\hline\hline
\end{tabular}
\label{TQQ'ss}
\end{table*}

\begin{figure}[h]
    \centering
    \includegraphics[width=8cm]{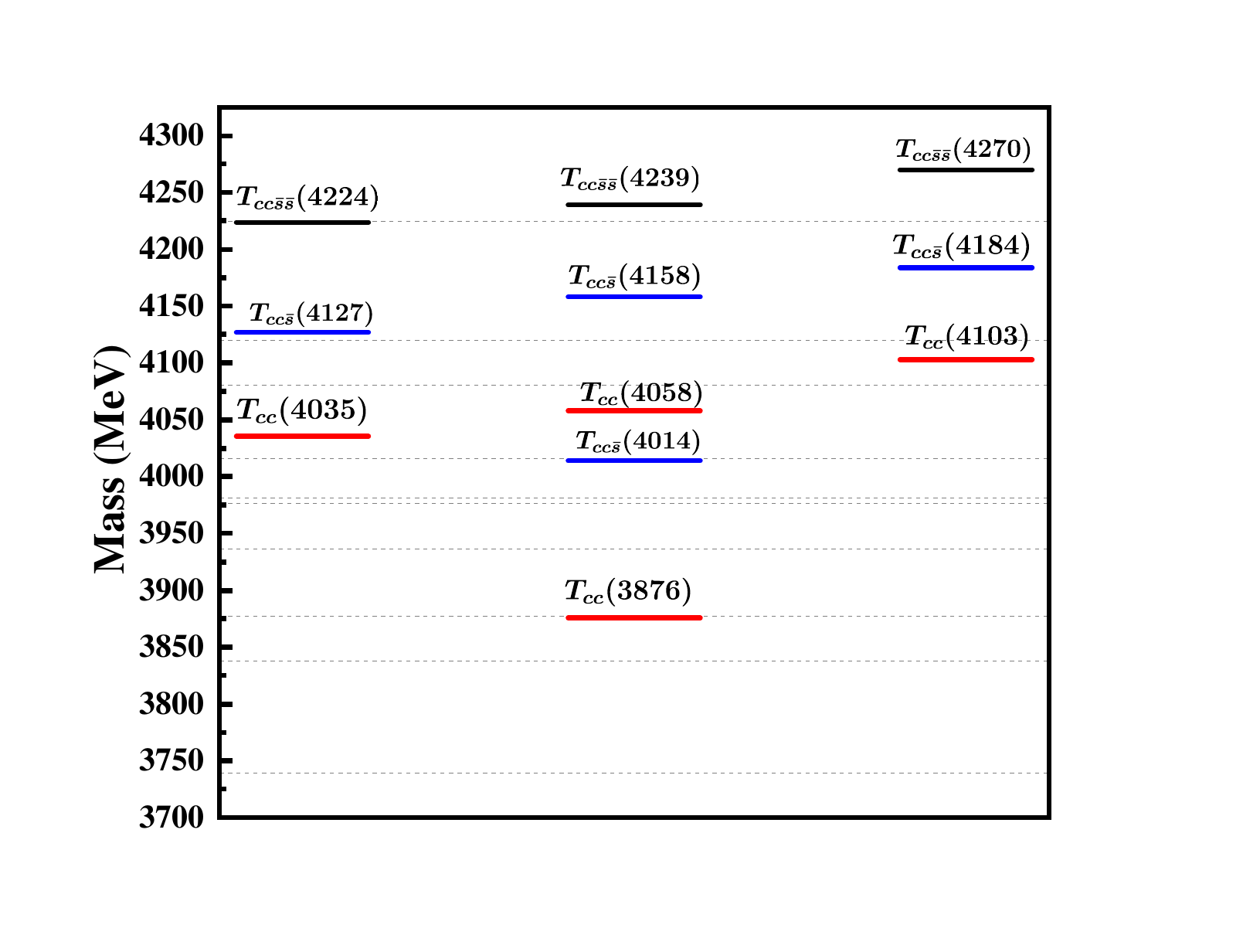}
    \caption{Mass spectrum and spin-parity of doubly charmed tetraquarks with nonstrange (red), single-strange (blue) and double-strange quarks (black).}
    \label{fig:Tcc}
\end{figure}

\begin{figure}[h]
    \centering
    \includegraphics[width=8cm]{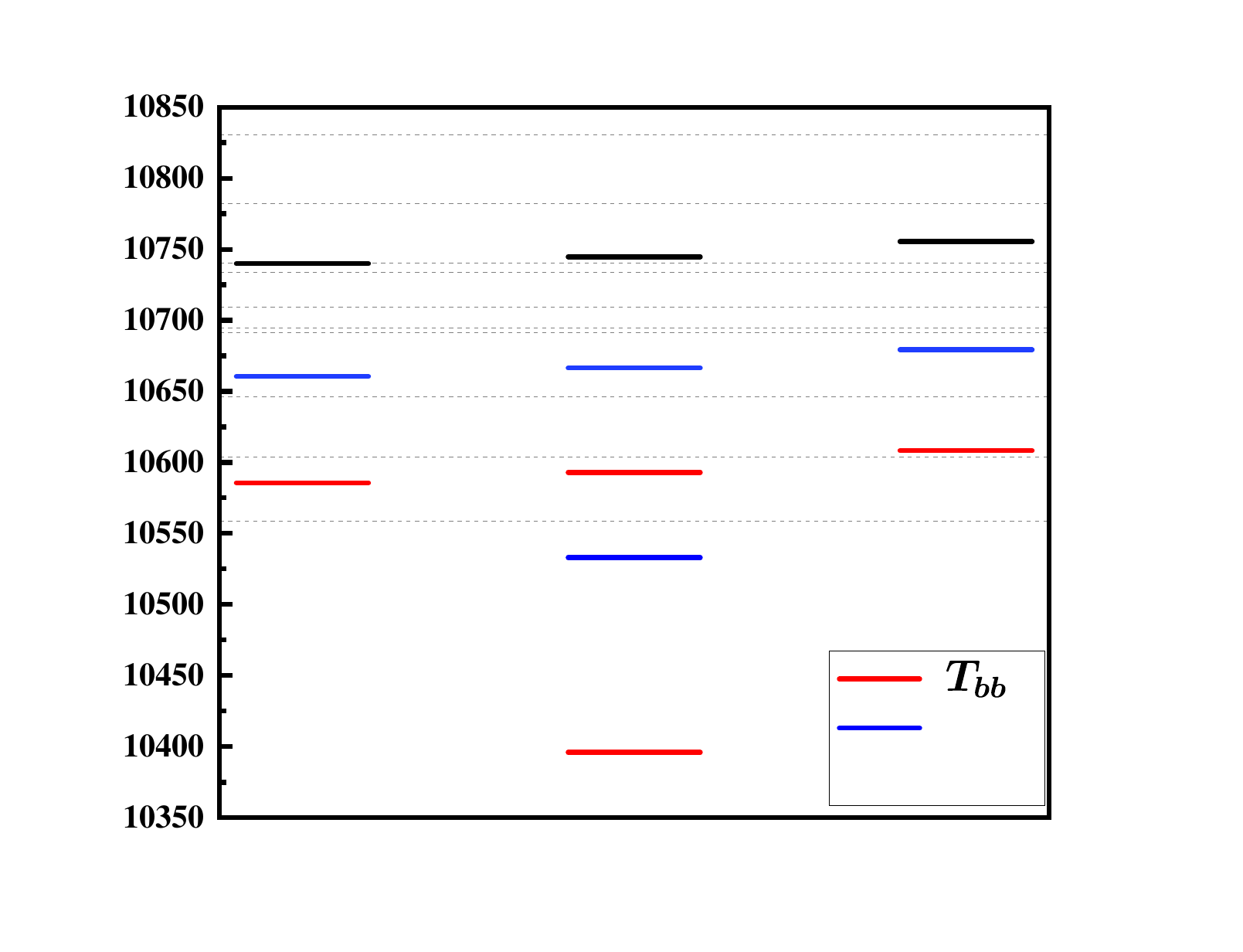}
    \caption{Mass spectrum and spin-parity of doubly bottomed tetraquarks with nonstrange (red), single-strange (blue) and double-strange quarks (black).}
    \label{fig:Tbb}
\end{figure}

\begin{figure}[h]
    \centering
    \includegraphics[width=8cm]{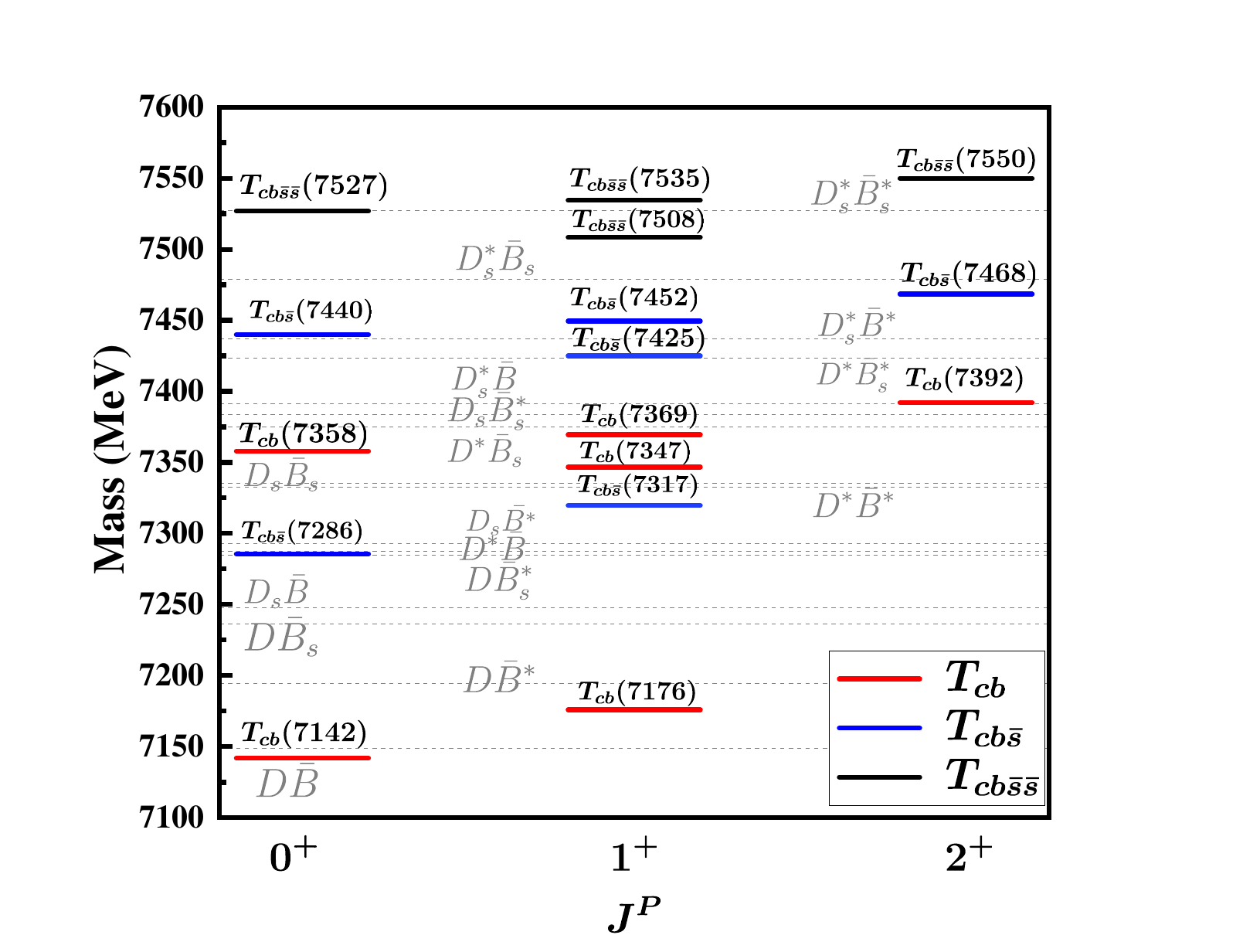}
    \caption{Mass spectrum and spin-parity of bottom-charmed tetraquarks with nonstrange (red), single-strange (blue) and double-strange quarks (black).}
    \label{fig:Tcb}
\end{figure}

The mass spectra of DHTs are presented in Figs.~\ref{fig:Tcc},~\ref{fig:Tbb} and ~\ref{fig:Tcb}, which reveal systematic patterns across different flavor sectors. All figures share common features: states are categorized by $J^P$ quantum numbers ($0^+$, $1^+$, $2^+$) and color-coded by strangeness content (non-strange: red, single-strange: blue, double-strange: black), with dashed lines indicating relevant meson-meson thresholds.

Fig.~\ref{fig:Tcc} displays the doubly charmed tetraquark ($T_{cc\bar{q}\bar{q}'}$) spectrum, there are 11 particles in total, consisting of 3 $T_{cc\bar{s}\bar{s}}$ states, 4 $T_{cc\bar{s}}$ states and 5 $T_{cc}$ states. 
Notably, the $T_{cc}(3876)$ state with $J^P=1^+$ aligns closely with the experimentally observed $T_{cc}^+(3875)$, positioned near the $D^{*+}D^0$ threshold. 
The $T_{cc\bar{s}}(4014)$ mass is above the $DD^*_s$ and $D^*D_s$ thresholds, but below that of the $D^*D^*_s$, while other $T_{cc\bar{q}\bar{q}'}$ states lie above relevant meson-pair thresholds. 
Fig.~\ref{fig:Tbb} shows the mass spectrum of ground doubly bottomed heavy tetraquarks $T_{bb\bar{q}\bar{q}'}$. The numbers of $T_{bb\bar{q}\bar{q}'}$ in Fig.~\ref{fig:Tbb}
are consistent with that in Fig.~\ref{fig:Tcc} because they share the same configurations.  However, unlike the ground doubly charmed heavy tetraquarks, there are many states whose masses are below the corresponding meson pair thresholds. All the states in the $1^+$ and $2^+$ configurations are below the corresponding meson pair masses, particularly the $T_{bb}(10396)$ with $J^P=1^+$ which sits $\sim$ 160 MeV beneath the $B^*B$ threshold, suggesting exceptional stability. 
While in the $0^+$ configuration, the $T_{bb\bar{s}\bar{s}}(10740)$ is slightly higher than $\bar{B}_s\bar{B}_s$ threshold, $T_{bb\bar{s}}(10660)$ is higher than $\bar{B}\bar{B}_s$, and $T_{bb}(10586)$ is higher than $\bar{B}\bar{B}$, respectively. 
Fig .~\ref{fig:Tcb} shows the bottom-charm $T_{bc\bar{q}\bar{q}'}$ sector featuring 13 states due to asymmetric spin couplings.
Among them, there are 4 $T_{cb\bar{s}\bar{s}}$ states, 6 $T_{cb\bar{s}}$ states, and 6 $T_{cb}$ states. 
 Among the 5 states in $0^+$ configuration, $T_{cb\bar{s}}(7286)$ is higher than $D\bar{B}_s$ and $D_s\bar{B}$, but much lower than $D^*_s\bar{B}^*$ and $D^*\bar{B}^*_s$,
$T_{cb}(7142)$ is below the $D\bar{B}$ threshold, while the remaining states are above the corresponding meson pair thresholds. 
In the $1^+$ and $2^+$configuration, most of the  $T_{bc\bar{q}\bar{q}'}$ states are above the corresponding meson-meson thresholds, except the $T_{cb\bar{s}}(7317)$, which is above the threshold of $D\bar{B}^*_s$ but below that of $D^*\bar{B}_s$.

Common characteristics include the universal mass hierarchy $T_{QQ'} < T_{QQ'\bar{s}} < T_{QQ'\bar{s}\bar{s}}$, reduced hyperfine splittings in heavier flavor sectors, and enhanced binding for $b$-quark systems due to diminished kinetic energy.
The progressive stabilization from $cc$ to $bb$ systems highlights the mass-dependence of binding mechanisms. 
Furthermore, by systematically comparing the obtained mass spectrum of DHTs with the corresponding meson-meson thresholds, we find that most of the DHT masses are above or far from the thresholds. Considering that molecular states are generally close to and below the thresholds, this provides crucial diagnostics for distinguishing compact tetraquarks from corresponding molecular states.

\section{Summary}
In this work, we provides a comprehensive investigation of DHTs ($T_{QQ\bar{q}\bar{q}'}$) spectrum using HADS within a well accepted constituent quark model framework calibrated to established hadron data. By correlating $QQ\bar{q}\bar{q}'$ states with experimentally measured heavy baryon ($Qqq'$) spectrum through HADS, we systematically predict masses for 38 ground-state tetraquarks with $J^P = 0^+,1^+,2^+$ across $cc$, $bb$, and $bc$ flavor sectors with varying strange quark content. All parameters derive from baryon spectra fits, ensuring model consistency with QCD phenomenology without new inputs. 
Key findings reveal that $bb$-containing tetraquarks predominantly lie below meson-meson thresholds, indicating strong binding and potential stability, while $cc$ and $bc$ configurations generally reside near or above thresholds except for specific cases like the $T_{cc}(3876)$ state that matches the observed $T_{cc}^+(3875)$. The universal mass ordering $T_{QQ'} < T_{QQ'\bar{s}} < T_{QQ'\bar{s}\bar{s}}$ and method-independent hyperfine splittings provide critical discriminants between compact tetraquarks and molecular interpretations. 

Our study shows that the hyperfine splitting of DHT states decreases with the increase in the mass of heavy diquarks, which is consistent with the requirements of heavy quark symmetry. We have also estimated the impact of HADS symmetry breaking on the mass of DHTs, and found that for double-charmed, bottom-charmed, and double-bottomed tetraquark states, the impact levels are about 11.3 MeV, 5.7 MeV, and 3.8 MeV respectively. This order relationship also reflects the heavy quark mass dependence of heavy quark symmetry.
These predictions establish essential benchmarks for experimental searches at LHCb, Belle II, and PANDA, while demonstrating HADS as a fundamental symmetry unifying conventional and exotic heavy-quark hadrons.

\section{Acknowledgments}The work of T.~W.~W is supported by the National Natural Science Foundation of China under Grant No.12405108. Y.~L.~M. is supported in part by the National Science Foundation of China (NSFC) under Grant No. 12347103 and Gusu Talent Innovation Program under Grant No.
ZXL2024363.

    \bibliography{doubly-heavy}
\end{document}